\documentclass[aps,preprint]{revtex4}
\usepackage{graphicx}

\begin{document}

\title{Investigation of a protein complex network}

\author{A. R. Mashaghi}
 \email{mashaghi@ibb.ut.ac.ir}
 \address{Institute of Biochemistry and Biophysics , P.O.Box 13145-1384, Tehran, Iran}
 \address{School of Medicine, Tehran University, P.O.Box 14155-6447,Tehran,Iran}

\author{A. Ramezanpour}
 \email{ramzanpour@mehr.sharif.edu}
 \address{Department of Physics, Sharif University of Technology, P.O.Box
11365-9161, Tehran, Iran}

\author{V. Karimipour}
 \email{vahid@sharif.edu}
 \address{Department of Physics, Sharif University of Technology, P.O.Box
11365-9161, Tehran, Iran}

\begin{abstract}
The budding yeast {\it Saccharomyces cerevisiae} is the first
eukaryote whose genome has been completely sequenced. It is also
the first eukaryotic cell whose proteome (the set of all proteins)
and interactome (the network of all mutual interactions between
proteins) has been analyzed. In this paper we study the structure
of the yeast protein complex network in which weighted edges
between complexes represent the number of shared proteins. It is
found that the network of protein complexes is a small world
network with scale free behavior for many of its distributions.
However we find that there are no strong correlations between the
weights and degrees of neighboring complexes. To reveal non-random
features of the network we also compare it with a null model in
which the complexes randomly select their proteins. Finally we
propose a simple evolutionary model based on duplication and
divergence of proteins.
\end{abstract}
\maketitle

\section{Introduction}\label{1}
In recent years complex networks have attracted much of interest
to model real-life networks such as social, biological and
communication networks \cite{ab,dm,n1}. Certainly, introducing the
essential static and dynamic features of these networks can help
us in a better understanding of their various properties
\cite{ws,bw,asb,m,ajb,cnsw}. A well known property of most of
these networks, the so called small world property \cite{ws}
indicates that the average distance between any two nodes
increases slowly with the size of the network (i.e. as logarithm
of the size). This in turn can lead to a fast spreading of effects
in the network and so increases the finite size effects when one
studies for example diffusion in such a network \cite{aks}.
Extensive studies also indicate the importance of degree (the
number of neighbors of a node) distribution for static and dynamic
behaviors of the network \cite{ab,dm}. One can also add various
kinds of correlations, e.g. degree correlation of two neighbors,
to the list of these important features \cite{n2,n3}.\\
Protein interaction networks are important examples of the
real-life networks in which nodes and edges represent proteins and
interactions between them respectively \cite{u,i,a,w,pss,ms,vfmv}.
Proteins have been traditionally recognized on the basis of their
roles as enzymes, signalling molecules or structural components in
cells and micro-organisms. The most rudimentary structural
information about the proteome (assembly of proteins in an
organism) is the pattern of interactions between different
proteins. Determining such connections, helps us in understanding
the backbone of functional relationships between proteins and the
pathways for the propagation of various signals among them.
Besides specific detailed information about the pattern of
interactions in a single proteome, which are certainly important
for its functioning, some general characteristics of these
networks are also important in that they may point to universal
properties of organisms. For example, it has been shown that as
far as the interaction of individual proteins are concerned, the
interaction network of the budding yeast {\it Saccharomyses
cerevisiae} ({\it S. cerevisiae}) is a scale free network. This
property is itself a hint to the robustness of the protein
interaction network against the random removal of proteins
\cite{jmbo,ajb}.\\
However recent progress indicates that each of the central
processes in a cell \textit{is catalyzed not by a single protein
but by the coordinated action of a highly linked set of several
proteins, called a protein complex} \cite{ks,g,h}. Thus complexes
act as protein machines and have been evolved for the same reason
that humans have invented mechanical and electronic machines. It
is also remarkable that a single protein may be shared in several
different complexes. It is expected that a protein with a general
functionality is shared by many complexes. On the other hand it
seems that the weight of a complex depends on the degree of
sophistication of its tasks.\\
It is in view of this new emerging picture of the proteome as a
coordinated ensemble of protein complexes that we study the
general properties of the network of protein complexes of the
budding yeast. Certainly, analysis of the proteome map at both
protein and complex level will result in a better understanding of
the proteom functioning. Although from a biological point of view
the precise nature of proteins and their interactions in a single
living organism are important, from a physical point of view in
which we seek the general universal patterns among many different
organism \cite{p}, we can study the most elementary features of a
proteome, i.e. the weight distribution of complexes, distribution
of the number of proteins
shared between two complexes, etc.\\
Based on extensive information provided in \cite{g} we constructed
a weighted graph corresponding to the network of complexes. By
this we mean that both the nodes and the edges are assigned
weights. The weight of a node is equal to the number of proteins
in the complex it represents.  Two nodes are connected by a
weighted edge, whose weight indicates how many proteins are common
in the two complexes. Such a point of view seems to be the first
step to understand the integration and coordination of cellular
functions. Connections in this network not only reflect physical
interaction of complexes, but may also represent common
regulation, localization, turnover or architecture \cite{g}. In
addition it has a meaningful interpretation in other real-life
networks such as social networks when a connection between two
communities can only be established by common individuals in them.
Note that one may consider a single protein complex as a subgraph
of the protein interaction network with a high level of
interconnection between its elements. From this point of view the
protein complex network is a large scale or coarse grained picture
of the protein interaction network. However, we stress that in
view of the recent findings, this is not the correct picture for
the interactome. In fact the important feature of the recent
experiments \cite{g,h} compared to the previous experiments
\cite{u,i} is that they uncover not only the pairwise interactions
but ternary, quaternary and higher interactions between different
proteins in a complex
\cite{e}.\\
In this work we show that the above protein complex network is a
small world one with scale free degree distribution. Moreover it
is found that distributions of weights of complexes, weights of
edges and coordination numbers of proteins (the number of
complexes a protein participates) follow power law behaviors which
can in turn refer to a kind of preferential attachment in the
evolution of the protein complex network \cite{ab}. We compare the
network with some null models such as Erd\"{o}s and R\'{e}nyi
random graph \cite{er} and a random-selection model in which each
complex selects its proteins randomly from the list of all
proteins. Unlike the former case the latter model results in a
considerable clustering for the network of complexes. Here by
clustering we mean the probability that two neighbors of a node be
also connected to each other. In this manner the clustering is
equivalent to the transitivity of the network (defined as three
times the ratio of the number of triangles in the network to the
number of connected triples of nodes \cite{n4}). Moreover, we find
that the random-selection model can well reproduce the degree
distribution of complexes and the dependence of degree of a
complex on its weight. However some important distributions of the
network such as weight of edges are still far from the predictions
of this model in the region of large weights. Following the
previous models for the evolution of the protein interaction
network \cite{w,vfmv,pss}, we propose a simple evolutionary model
based on duplication and divergence (mutation) of proteins. We
show that this model reproduces the power law behavior of some
essential distributions such as weight of complexes and weight of
edges.

The paper is organized as follows. Section (\ref{2}) is devoted to
the description of the budding yeast protein complex network based
on the data provided in \cite{g}. Section (\ref{3}) gives a
comparison between the random-selection model and the real
network. The evolutionary model is introduced in section
(\ref{4}). We conclude the paper in section (\ref{5}).

\section{structural properties of the yeast protein complex network}\label{2}
\begin{figure}
  \centering
\includegraphics[width=8cm]{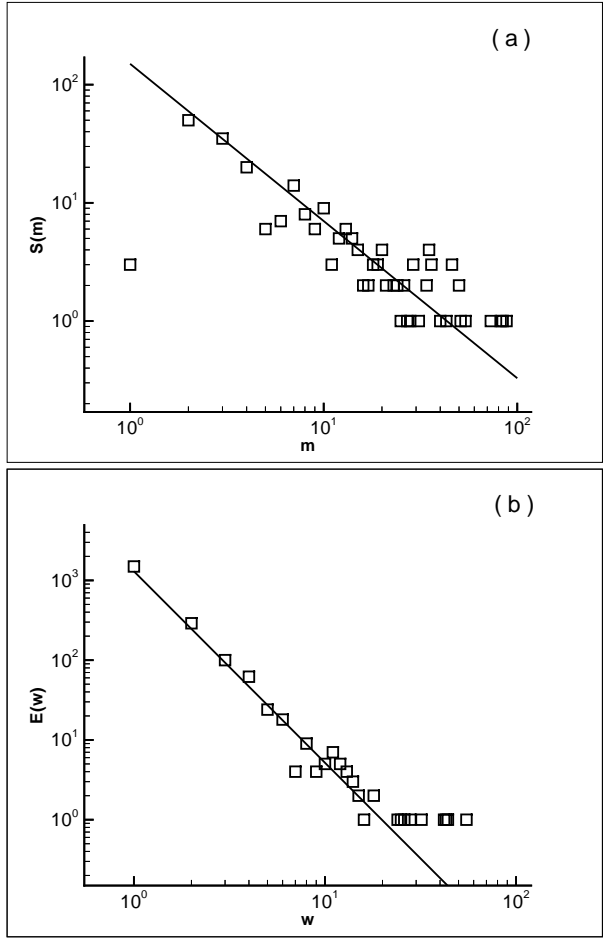}
    \caption{(a) Weight distribution of complexes and (b) weight distribution
    of edges in the yeast protein complex network. Lines in both figures
    show fitted power laws with the exponents given in the text.}\label{f1}
\end{figure}

According to the data we have extracted from \cite{g}, the {\it S.
cerevisiae} includes $n=1398$ proteins, organized in $N=232$
complexes. The resulting network is then composed of $232$ nodes
and $E=2043$ edges. A complex of weight $m$ is denoted by a node
of weight $m$, and if two complexes share $w$ proteins, a weight
$w$ is assigned to the edge connecting them. Figures (\ref{f1}-a)
and (\ref{f1}-b) show the distribution of weights of the nodes,
$S(m)$, and weights of the edges, $E(w)$.
Both distributions show the power law behavior ($S(m)\sim
m^{-\tau_m}$ and $E(w)\sim w^{-\tau_w}$) with exponents
respectively equal to $\tau_m=1.33 \pm 0.06$ and $\tau_w=2.4 \pm
0.1$, figures (\ref{f1}-a and \ref{f1}-b). The observed deviation
in the number of complexes of weight $1$ from the expected power
law behavior could be mostly attributed to the experimental
limitations \cite{g}. The average weight of the nodes and the
edges turn out to be respectively $\overline{m}=11.48$ and
$\overline{w}=1.79$ with the following dispersions: $\sigma_m
:=\sqrt{\overline{m^2}-\overline{m}^2}=14.69$ and $\sigma_w=2.9$.
One could attribute the large values of these dispersions to the
small size of the network and scale free nature of the related
distributions. A curious property of the network is that there is
no correlation between the weights of adjacent nodes. In other
words, the probability that an emanating edge from a complex of
weight $m$ encounters another complex of weight $m'$ is
independent of $m$. To show this we compute the associated
correlation coefficient \cite{n2}, $r_{mm}$, which in the protein
complex network has the value $-0.004$ and is defined as follows:
let $P(m,m')$ be the probability that an arbitrary edge lies
between two complexes of weights $m$ and $m'$. Thus
$\Pi(m)=\sum_{m'}P(m,m')$ gives the probability of finding a
complex of weight $m$ at the end point of an arbitrary edge. Now
the correlation coefficient is given by
\begin{equation}\label{rmm}
  r_{mm}:=\frac{\sum_{m,m'}mm'(P(m,m')-\Pi(m)\Pi(m'))}{\sum_m m^2\Pi(m)- (\sum_m m\Pi(m))^2}.
\end{equation}
It is clear that in the absence of any correlation (i.e. when
$P(m,m')=\Pi(m)\Pi(m')$) we have $r_{mm}=0$. If similar complexes
have a high tendency for being connected to each other we say that
the network is assortative \cite{n2} in this respect and the
correlation coefficient will be positive. Otherwise the network is
dissortative and the correlation coefficient will be negative.
Obviously one can apply this definition to any two point
distribution to measure the degree of correlation between the two
variables.\\

\begin{figure}
  \centering
\includegraphics[width=8cm]{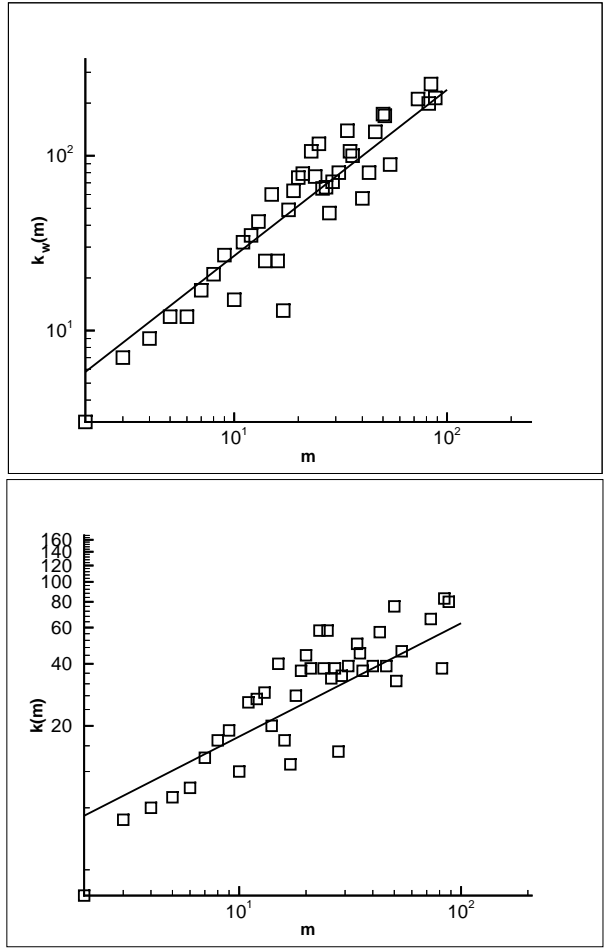}
    \caption{Dependence of (a) weighted degree and (b)
     degree of a complex on its weight.}\label{f2}
\end{figure}

One may also ask what is the relation between the degree of a
complex and its weight. Our results again give a power law
dependence for large values of complex weights. The total weight
of edges emanating from a complex of weight $m$ (its weighted
degree) scales as $k_w(m)\propto m^{\beta_w}$, where $\beta_w
\simeq 0.95 \pm 0.07$, figure (\ref{f2}-a).
It is found that the number of neighbors behaves in a similar way
$k(m)\propto m^\beta$ with $\beta\simeq 0.55 \pm 0.07$, figure
(\ref{f2}-b). Obviously $k_w(m)$ grows faster than $k(m)$ with the
weight of complex as expected. Roughly speaking, since $\beta_w
\approx 2\beta$, the above relations suggest that the average
weight of an edge
emanating from a complex of weight $m$ scales as $m^{\beta}$.\\

\begin{figure}
  \centering
\includegraphics[width=8cm]{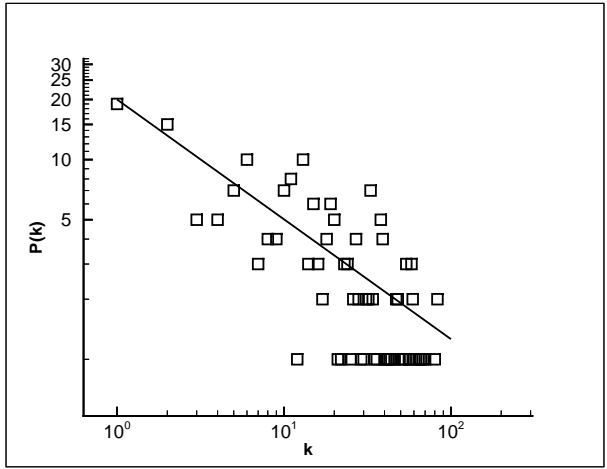}
    \caption{ Degree distribution of complexes.}\label{f3}
\end{figure}

\begin{figure}
  \centering
\includegraphics[width=8cm]{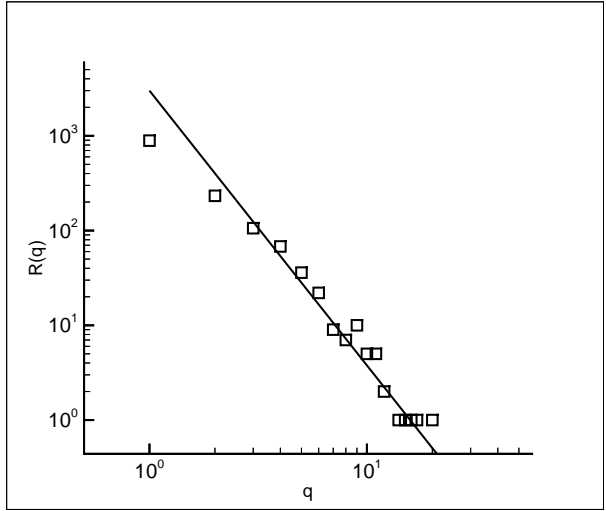}
    \caption{Coordination number distribution of proteins in the
     yeast protein complex network.}\label{f4}
\end{figure}

To address topological properties of the graph we studied also the
degree distribution, $P(k)$ and the correlation between degrees of
neighboring nodes which can again be detected by computing the
related correlation coefficient, $r_{kk}$. Note that by degree of
a node we mean the number of neighboring nodes, irrespective of
the weights of the edges emanating from that node. By weighted
degree of a node we mean the sum of weights of the edges emanating
from this node. In figure (\ref{f3}) we show that the degree
distribution is scale free with a sharp cutoff at its tail. In
this figure we see also a power law fit of form $P(k)\sim
k^{-\gamma}$ with exponent $\gamma=0.6 \pm 0.04$ to the real data.
Moreover, we find the value $-0.06$ for $r_{kk}$ which means that
in contrast to the protein interaction networks \cite{ms}, there
is no strong correlation between the degrees of adjacent nodes in
protein complex network. Our results shows that the same
conclusion is true if we
take also the weights of the edges into account.\\
Usually a given protein takes part in more than one, say $q$
complexes. We call $q$ the coordination number of that protein. We
found that on average a protein takes part in $\overline{q}=1.91$
complexes (with $\sigma_q=1.86$). Figure (\ref{f4}) shows the
number of proteins versus their coordination numbers, $R(q)$. The
distribution is a scale free one for large coordination numbers,
that is $R(q)\sim q^{-\tau_q}$ with $\tau_q=2.95 \pm 0.12$ as its
exponent.
This indicates that there are a few number of proteins with high
coordination numbers. It seems that these are the proteins with
exceedingly important role in the functioning of the cell. Let us
compute the correlation coefficient which measures the correlation
between coordination of proteins and weight of complexes they
contribute. To this end we consider the collection of proteins and
complexes as a bipartite network in which nodes of the first type
represent complexes and those of the second type represent
proteins. An edge in this network connects a node of the first
kind to a node of the second kind. Thus the number of edges
emanating from a protein determines its coordination number and
similarly the number of edges connected to a complex gives its
weight. Now we define $P(m,q)$ as the probability that an
arbitrary edge of this bipartite network connects a complex of
weight $m$ to a protein of coordination number $q$. Therefor
$\Pi(m):=\sum_qP(m,q)$ and $\pi(q):=\sum_mP(m,q)$ are respectively
the probability that an arbitrary edge reaches to a complex of
weight $m$ and a protein of coordination number $q$. Now the
associated correlation coefficient is defined as
\begin{equation}\label{rmq}
  r_{mq}:=\frac{\sum_{m,q}mq(P(m,q)-\Pi(m)\pi(q))}
  {\sqrt{\sum_m m^2\Pi(m)- (\sum_m m\Pi(m))^2}
  \sqrt{\sum_q q^2\pi(q)- (\sum_q q\pi(q))^2}}.
\end{equation}

We find that in the the case of our data $r_{mq}=0.024$ which
indicates to the absence of correlation in this respect. That is
the coordination number of a protein does not affects in its
relation with complexes of different weights more than what is
expected by chance.\\

The network of complexes also defines pathways for the propagation
of various signals such as phosphorylation and allosteric
regulation of proteins. For such functions, a key parameter of the
network is its diameter, defined as the shortest path between the
remotest nodes in the giant component of the network. Our analysis
reveals that the network has a small diameter, $D=5$, which points
to the small world property of the network. To confirm this we
also computed $\overline{1/d_{i,j}}$ (where $d_{i,j}$ denotes the
shortest distance between nodes $i$ and $j$) and compared it with
the corresponding quantity in an equivalent Erd\"{o}s - R\'{e}nyi
random graph, table (\ref{t}). The reason behind the computation
of $\overline{1/d_{i,j}}$ is that the above graph is not connected
rather it has a giant component of size $S_g=198$, a binary
component and $32$ single nodes. To have a measure of clustering
in the network, its transitivity $T$ was extracted and it has been
found that it is almost six times greater than the one in an
Erd\"{o}s - R\'{e}nyi random graph, see table (\ref{t}). This high
level of transitivity which is another hint to the small world
property of the protein complex network would give rise to the
robustness of the network against the random removal of nodes or
edges \cite{n4}.

\section{The Random-Selection Model}\label{3}
\begin{figure}
  \centering
\includegraphics[width=8cm]{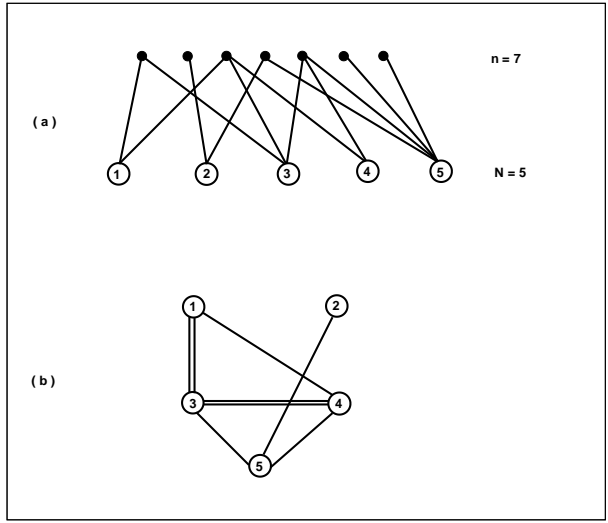}
    \caption{(a) A bipartite graph of $N$ complexes and $n$ proteins. Connections can only be
    established between proteins and complexes. (b) the resulted
    weighted graph of protein complexes. The number of lines connecting two complexes
    represents the number of shared proteins.}\label{f5}
\end{figure}

\begin{figure}
  \centering
\includegraphics[width=8cm]{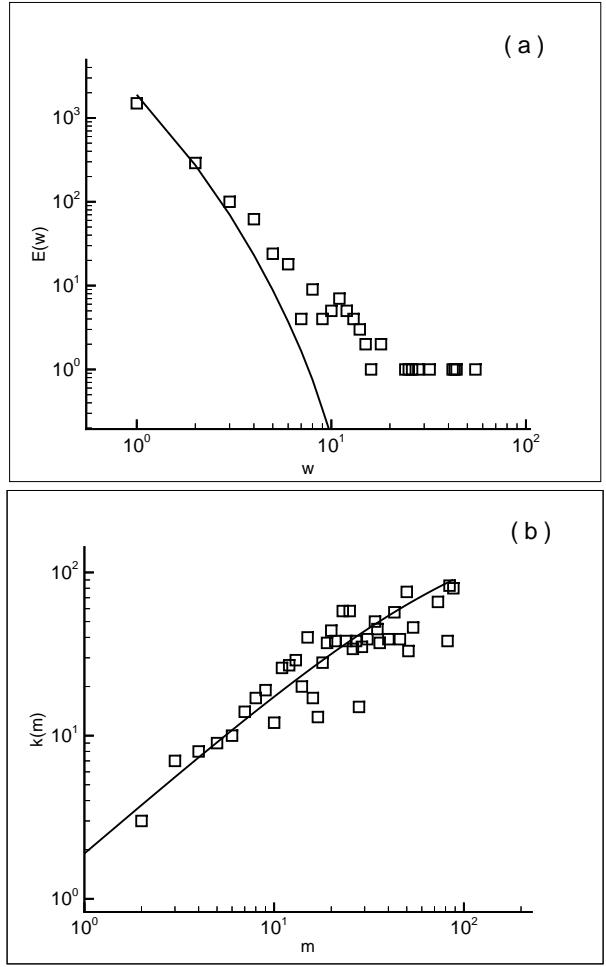}
    \caption{Comparison of the {\it S. cerevisiae} protein complex network (squares)
     with the analytic predictions of the random-selection model (lines): (a)
     weight distribution of edges and (b) degree of a complex versus its weight}\label{f6}
\end{figure}

To highlight the special features of the yeast protein complex
network and to possibly draw conclusions of biological interest,
one may compare it to a random network in which proteins aggregate
randomly to form different complexes. We call such a model a
random-selection model. The simplest such model may be constructed
as a bipartite network \cite{nsw} as follows: one takes a
bipartite network consisting of two types of nodes. Nodes of the
first type represent complexes and those of the second type
represent proteins. A bipartite network of this kind consisting of
$N$ complexes and $n$ proteins and the resulting weighted network
consisting of only protein complexes are depicted in figure
(\ref{f5}).
Here we start from a bipartite network and calculate many of the
properties of the resulting protein complex network exactly. A
given complex $\mu$ contains a number of proteins which we denote
by $m_{\mu}$. In general $ 0 < m_{\mu} \leq n$. From the real data
one can infer the actual sequence $(m_1, m_2, \cdots m_N)$.
 One thus assigns free (unconnected) stubs to
the first type of nodes (the complexes) according to this sequence
and then connects these stubs randomly to the second types of
nodes (proteins). Each protein connected to a complex means that
it is contained in that complex. In this way one obtains another
distribution $R(q)$, where $q$ is the number of complexes a given
protein participates in. Clearly a given protein may be contained
in more than one complex. Note that a peculiar feature of this
model is that certain proteins may not be connected to any complex
at all. Later in this section we introduce another
slightly improved model in which this effect does not happen. \\
In the following we calculate many of the interesting quantities
of the resulting weighted network of complexes exactly and compare
them with real data to check the viability of this model. To
proceed let $ S(m)$ denotes the number of complexes of weight $m$.
Obviously $\sum_m{S(m)}= N $, where $N$ is the total number of
complexes. First let us calculate the probability of two complexes
of weight $ m$ and $ m'$ to have $ w $ proteins in common. Denote
this probability by $ P(m,m';w)$. The first complex chooses its
$m$ members freely from the collection of all proteins. The second
complex, has $
 \left(%
\begin{array}{c}
  n \\
  m' \\
\end{array}%
\right)$ ways for choosing its members from the collection, of
which $ \left(%
\begin{array}{c}
  m \\
  w \\
\end{array}%
\right)$ ways are available for choosing $w$ members in common
from the first complex and $\left(%
\begin{array}{c}
  n-m \\
  m'-w \\
\end{array}%
\right)$ ways are available for choosing the remaining set
disjoint from the first complex. Hence the probability is
\begin{equation}\label{pww1}
P(m,m';w)=\frac{\left(%
\begin{array}{c}
  m \\
  w \\
\end{array}%
\right)\left(%
\begin{array}{c}
  n-m \\
  m'-w \\
\end{array}%
\right)}{\left(%
\begin{array}{c}
  n \\
  m' \\
\end{array}%
\right)}
\end{equation}.
This equation can be rewritten in the form
\begin{equation}\label{pww2}
  P(m,m';w)=
  \frac{m!m'!}{(m-w)!(m'-w)!}\frac{(n-m)!(n-m')!}{n!w!(n+w-m-m')!},
\end{equation}
to make its symmetry under interchange of $m$ and $m'$ manifest.
The probability that two such complexes have no common members (be
not connected to each other in the network) is:
\begin{equation}\label{pww0}
  P(m,m';0)=\frac{(n-m)!(n-m')!}{n!(n-m-m')!}.
\end{equation}
For $ m,m'<< n$ one can approximates $(n-m)!/n!$ by $n^{-m}$ and
$(n-m')!/(n-m-m')!$ by $(n-m')^m$. So the above relation takes the
form
\begin{equation}\label{pww01}
P(m,m';0)\simeq (1-\frac{m'}{n})^m \simeq e^{-mm'/n}.
\end{equation}
Thus the probability of two complexes being connected will be
$1-P(m,m';0)$ and the average number of edges will be:
\begin{equation}\label{E}
E=\frac{1}{2}\sum_{m,m'}{S(m)S(m')(1-P(m,m';0)}.
\end{equation}
Moreover the average number of edges with weight $w$ is given by:
\begin{equation}\label{Ew}
E(w)=\frac{1}{2}\sum_{m,m'}{S(m)S(m')P(m,m';w)}.
\end{equation}
Figure (\ref{f6}-a) shows this quantity for a model network with
the same parameters as the yeast protein complex network.
We see that the weight distribution of edges decreases
exponentially in contrast to the power law behavior of the protein
complex network. This means that in the yeast protein complex
there are edges with very high weights. These high weight edges
inevitably connect high weight complexes. Thus in the yeast
complex network, the hubs (complexes with high degree) are
connected to each other intensively. This is in contrast with the
protein interaction network \cite{ms} where
the high degree proteins have a larger tendency to be connected to low degree ones.\\
In the same way one obtains $k(m)$, the average number of
neighbors of a given complex of weight $m$:
\begin{equation}\label{km}
    k(m)=\sum_{m'}{S(m')(1-P(m,m';0))}
\end{equation}
In figure (\ref{f6}-b) we compare the above expression with the
one obtained from the real network. One finds a good agreement
between predictions of the model and the real data. From the
distribution of weights of complexes $S(m)$ and the above
relation, one can obtain the distribution of degrees $P(k)$ using
the relation $ P(k)\Delta k=S(m)\Delta m$. Thus we expect that
degree distribution of the random-selection model to be also close
to the real one. \\

What can be said about $ r(q):=\frac{R(q)}{n}$, the probability
that a given protein contributes in $q$ complexes? It is not
difficult to show that due to the fully random nature of the
selection process, the above distribution has a binomial form like
\begin{equation}\label{rk0}
r(q)=(\begin{array}{c}N\overline{m} \\ q \end{array}
)(\frac{1}{n})^q(1-\frac{1}{n})^{N\overline{m}-q}
\end{equation}
where $\overline{m}$ is the average weight of complexes. Thus for
very large values of $N$ and $n$ we have
\begin{equation}\label{rk}
r(q)=\frac{\lambda^q}{q!}e^{-\lambda}, \hskip 1cm
\lambda=\frac{N}{n}\overline{m}.
\end{equation}
Therefore we get a Poisson distribution for coordination numbers
of proteins in distinctive
contrast to the real distribution which is scale free.\\

\begin{figure}
  \centering
\includegraphics[width=8cm]{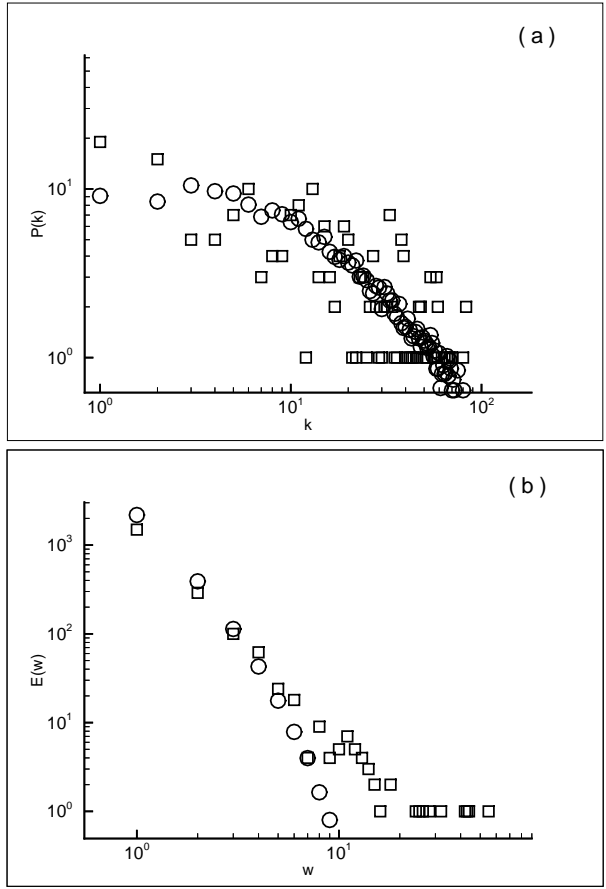}
    \caption{Comparison of the {\it S. cerevisiae} protein complex network (squares)
     with predictions of the improved random-selection model (circles): (a) degree distribution
     of complexes (b) weight distribution of edges}\label{f7}
\end{figure}

As seen from figure (\ref{f6}), our analytical treatment of this
model revealed that some of the general characteristics (the
degree of a given complex as a function of its weight and so the
degree distribution of complexes) of these networks are close to
those of {\it S. cerevisiae} protein complex network. Further
results derived from numerical simulations given in table
(\ref{t}) show that random-selection model is rather close to the
yeast protein complex network.\\
Note however that certain discrepancies between the
random-selection model and the real complex network, i.e. in the
distribution of the weights of edges, persist even if we improve
this model by fixing the coordination number of proteins from the
beginning to be the same as the one derived from real data of the
protein complex network (see figure (\ref{f7})). The evolutionary
model that we introduce in the next section is aimed to remove
this discrepancy.\\

\section{The Evolutionary Model}\label{4}
\begin{figure}
  \centering
\includegraphics[width=8cm]{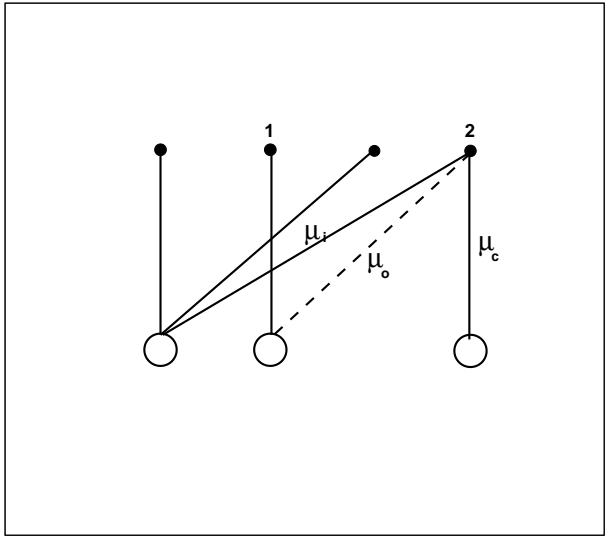}
    \caption{A typical step of the evolutionary model in which protein $1$
    has been duplicated and edges of the new protein
     (protein $2$) undergone mutations.}\label{f8}
\end{figure}

\begin{figure}
  \centering
\includegraphics[width=8cm]{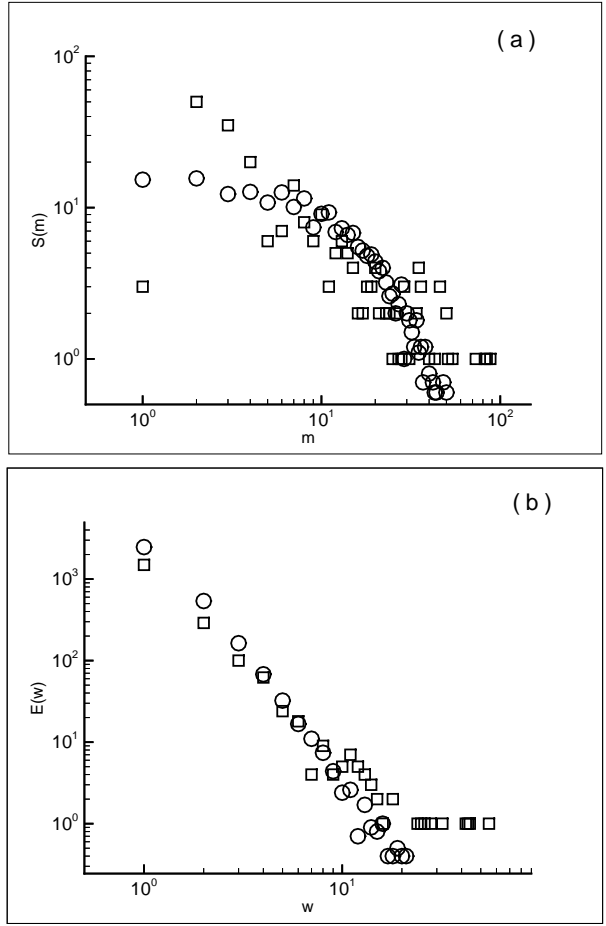}
    \caption{Comparison of the yeast protein complex network (squares) with the evolutionary model
    (circles):
     (a) weight distribution of complexes (b) weight distribution of edges
     The evolutionary model distributions are results of averaging over 2000
      runs of the evolutionary process.
      The associated relative statistical errors are in order of a few percent.}\label{f9}
\end{figure}

In this section we introduce a simple model aimed at representing
the evolution of protein complex network. The model is based on
the extension of the hypothesis of evolution by duplications and
divergence of proteins \cite{w,pss,vfmv} to the level of protein
complexes. DNA duplication has long been known as an important
factor in the evolution of genome size. This process almost
certainly explains the presence of large families of genes with
related functions in biological complex organisms. Consider a set
of proteins which belong to a number of functional units or
complexes. In each evolutionary step, the following actions take
place. A protein is randomly chosen and duplicated. It means that
a new protein is added to the proteome and contributes to all the
complexes in which its mother participates. Then the new protein
undergoes mutations; it loses its membership in any given complex
with probability $\mu_o$, and enters any other complex with
probability $\mu_i$. It is also probable that the protein creates
a new complex (novel functionality) by its own with probability
$\mu_c$. In figure (\ref{f8}) we have shown what happens in a
typical step of network evolution.
Note that during this process the new protein may not contribute
to any complex and thus leaves the proteome. Starting from one
protein in one complex, we have taken $t=1500$ evolutionary steps.
We found that it was the best number of steps which could
intimately produce the desired results. We found that in this
situation only about $7$ percent of duplicated proteins are stuff
and the others takes part at least in one complex. Having $t$ we
will also have the probability of creating a new complex in each
step, because $N$ the number of complexes satisfies $N=1+\mu_c t$.
The other parameters can also be determined by requesting that the
number of proteins and some important distributions (e.g. weight
distribution of complexes) be as close as possible to the real
data. Moreover, note that we also expect that the probability of
entering a new complex for a duplicated protein must be much
smaller than the probability of exiting one of its inherited
complexes. Summing up these points we find, by fitting to the real
data, the following values for parameters of the model: $\mu_o
\simeq 0.4$ , $\mu_i \simeq 0.01$ and $\mu_c \simeq 0.154$. One
can find results of this model in
figure (\ref{f9}) and table (\ref{t}).\\
In figure (\ref{f9}-a) we see that the model closely reproduces
the scale free behavior of the weight distribution of complexes in
the region of large weights although there are some deviations
from the real data for the number of small complexes. Indeed we
expected such a power law behavior in advance due to the presence
of preferential attachment\cite{ab} in the process of evolution of
complexes. Note that although the selection of proteins for
duplication is a completely random procedure, complexes with
higher weights have a higher chance to include the new duplicated
protein. In this way the evolution of complexes is governed by the
well known mechanism of preferential attachment which is a
principal way to produce scale free behaviors in the
realm of complex networks\cite{ab}.\\
Figure (\ref{f9}-b) shows the weight distribution of edges in the
resulted protein complex network of the evolutionary model. The
agreement with the real data is excellent. It is one of the
essential features of protein complex network which none of the
random models studied in this paper could exhibit it. We believe
that this fact indicates to the essential role of duplication and
divergence of proteins in the evolution of the protein complex
network.\\
Moreover as table (\ref{t}) shows the evolutionary model results
in a small world protein complex network. Both the transitivity
and diameter of the model are comparable with the values of the
real network of complexes.\\
This model also exhibits a little tendency for being a negative
correlated network in weights and degrees of neighboring
complexes. Indeed the related correlation coefficients turn out to
be $r_{mm}=-0.05 \pm 0.001$ and $r_{kk}=-0.08\pm 0.002$. It means
that complexes with different weights and degrees have a
larger probability to be connected to each other.\\
Finally it is worthwhile to note that the data studied in this
paper are those of the yeast proteome whereas the results given
in this section are the average behavior of such a proteome.

\section{Conclusion}\label{5}
To summarize, we have shown that the protein complex network of
the yeast is a nearly uncorrelated small world scale free network.
The power law behavior was also found in other significant
distributions such as weight of complexes, weight of edges and
coordination number of proteins. Although some of these features
(e.g. small world property and high clustering) were expected in
advance, this study revealed some distinctive properties of this
network like the scale free behavior of weight distribution of
edges. We also compared the yeast protein complex network with a
random-selection model and a simple evolutionary model. It was
found that the random-selection model can satisfactorily reproduce
the relation between weight and degree of a complex and also
degree distribution of the real network. However this model failed
to give the power law behavior of the distribution of weights of
edges, a property which could be well reproduced in the
evolutionary model. In the latter model the desired distributions
automatically arise
by just fitting the model parameters using the real data.\\
From the evolutionary point of view, the above study can also be a
hint to the essential role of duplication and divergence processes
in the evolution of proteome and this study is indeed an extension
of this mechanism to the level of protein complexes. Certainly
results of the previous studies on protein interaction network
along with the investigation of large scale properties in the
protein complex network will help in a better understanding of the
general behaviors of such systems.\\

\begin{table}

\begin{center}

\begin{tabular}{|c|c|c|c|c|c|}
  \hline

                         &Yeast PCN      & ER Model         & RS Model        & Improved RS Model& Evolutionary Model \\
  \hline
  N                      & 232           & 232              & 232             & 232              & 232.7$\pm$0.3 \\
  n                      & 1398          & ---              & 1398            & 1398             & 1364.7$\pm$0.4 \\
  $\overline{k}$         & 17.61 (18.95) & 17.6             & 17.5$\pm$0.1    & 23.8$\pm$0.05    & 29.57$\pm$0.06 \\
  $\overline{k_w}$       & 31.62 (45.73) & ---              & 21.8$\pm$0.1    & 31.62            & 43.6$\pm$0.08 \\
  $\overline{q}$         & 1.91 (1.86)   & ---              & 2.24$\pm10^{-2}$& 1.91             & 2.73$\pm3*10^{-3}$ \\
  T                      & 0.41          & 0.07$\pm 10^{-3}$& 0.29$\pm10^{-3}$& 0.39$\pm10^{-3}$ & 0.29$\pm10^{-3}$ \\
  D                      & 5             & 3.08$\pm 0.03$   & 4.4$\pm$0.1     & 4.4$\pm$0.1      & 4.02$\pm4*10^{-3}$ \\
  $\overline{\frac{1}{d_{i,j}}}$ & 0.35(0.27)   & 0.49$\pm 10^{-3}$& 0.47$\pm10^{-3}$& 0.48$\pm10^{-3}$ & 0.54$\pm10^{-4}$ \\
  \hline

\end{tabular}

\vskip 0.5cm

\caption{Comparison of the yeast protein complex network (PCN)
with some models: Erd\"{o}s and R\'{e}nyi (ER) random graph, the
random-selection model (RS), improved random-selection model and
the evolutionary model. The first column represents: number of
complexes (N), number of proteins (n), average of degree
($\overline{k}$), weighted degree ($\overline{k_w}$), coordination
number of proteins ($\overline{q}$), transitivity (T), diameter of
network (D) and inverse of shortest distance between two arbitrary
nodes ($\overline{1/d_{i,j}}$). Values in parenthesis refer to
dispersion of associated quantity. Statistical errors have been
denoted where we have done averaging over different
realizations.}\label{t}
\end{center}
\end{table}

\end{document}